\documentclass[conference]{IEEEtran}
\usepackage[utf8]{inputenc}
\usepackage{amsmath,amssymb,amsfonts}
\usepackage[absolute,overlay]{textpos}
\usepackage{textcomp}
\usepackage{algorithmic}
\usepackage{algorithm}
\usepackage{array}
\usepackage[caption=false,font=footnotesize]{subfig}
\usepackage{float}
\usepackage{stfloats}
\usepackage{url}
\usepackage{verbatim}
\usepackage{graphicx}
\usepackage{caption}
\usepackage{romannum}
\usepackage{cite}
\usepackage{makecell}
\usepackage{multirow}
\usepackage{xcolor}
\usepackage{adjustbox}
\usepackage{hyperref}
\usepackage{booktabs}   
\usepackage{cleveref}
\usepackage{titlesec}
\titlespacing*{\section} {0pt}{1ex}{1ex}
\titlespacing*{\subsection} {0pt}{0.5ex}{0.5ex}
\def\BibTeX{{\rm B\kern-.05em{\sc i\kern-.025em b}\kern-.08em
      T\kern-.1667em\lower.7ex\hbox{E}\kern-.125emX}}

\title{Utilizing Adversarial Training for Robust Voltage Control: An Adaptive Deep Reinforcement Learning Method

\author{
 \IEEEauthorblockN{Sungjoo Chung, Ying Zhang}
    \IEEEauthorblockA{\textit{School of Electrical and Computer Engineering} \\
    \textit{Oklahoma State University}\\
    Stillwater, U.S. \\
    \{sungjoo.chung; y.zhang\}@okstate.edu}
}

}

\begin{document}

\maketitle


\begin{abstract}
Adversarial training is a defense method that trains machine learning models on intentionally perturbed attack inputs, so they learn to be robust against adversarial examples. This paper develops a robust voltage control framework for distribution networks with high penetration of distributed energy resources (DERs). Conventional voltage control methods are vulnerable to strategic cyber attacks, as they typically consider only random or black-box perturbations. To address this, we formulate white-box adversarial attacks using Projected Gradient Descent (PGD) and train a deep reinforcement learning (DRL) agent adversarially.  The resulting policy adapts in real time to high-impact, strategically optimized perturbations. Simulations on DER-rich networks show that the approach maintains voltage stability and operational efficiency under realistic attack scenarios, highlighting the effectiveness of gradient-based adversarial DRL in enhancing robustness and adaptability in modern distribution system control.
\end{abstract}

\section{Introduction}

\IEEEPARstart{D}{riven} by the pursuit of energy efficiency and environmental sustainability, distributed energy resources (DERs) are being increasingly integrated into distribution networks \cite{Detchon}. Leveraging the potential of DERs to enhance grid reliability and flexibility requires secure and dependable coordination among control devices, communication infrastructure, and control centers. However, the extensive data exchange, both among field devices and between these devices and control centers, introduces significant cybersecurity vulnerabilities \cite{Tatiparti10418207}. These vulnerabilities can compromise the effectiveness of coordination strategies implemented by distribution management systems (DMSs), such as Volt-Var optimization (VVO), ultimately resulting in degraded reliability and operational efficiency of the distribution network.

VVO represents a core operational function of modern DMS, aiming to improve network operational efficiency and minimize reactive power losses by computing optimal control set-points for voltage-regulating devices, such as smart inverters (SIs), capacitor banks (CBs), on-load tap changers (OLTCs), and battery energy storage systems (BESSs) \cite{Satsangi7853324}. As VVO algorithms are critically dependent on accurate network state estimation, ensuring the integrity and security of measurement data and communication channels is essential. Consequently, robust cybersecurity mechanisms are imperative to safeguard the reliability and effectiveness of VVO deployment in distribution systems.

Motivated by these cybersecurity challenges, recent research has increasingly focused on enhancing the robustness of voltage control mechanisms under adversarial conditions. 
In \cite{Isozaki7112640}, a detection framework based on spatiotemporal consistency checks is developed to identify falsified voltage measurements in power systems with high photovoltaic (PV) penetration. A game-theoretic framework is presented in \cite{Teixeira6859265} to model the strategic interaction between an attacker injecting stealthy data and an operator defending an integrated Volt-Var control (VVC) system. 
A centralized VVO strategy is proposed in \cite{Majumdar7959131}, which computes reliable control set-points using historical and forecasted data and activates a Local Setting Solution under compromised conditions. In contrast to centralized and model-driven architectures, a fully autonomous VVC scheme is proposed in \cite{Joseph8954624} for DER-rich distribution networks, coordinating both static and dynamic reactive power control modes independently of external communication infrastructure. While these contributions have collectively advanced the state of secure voltage control, key limitations remain. Many existing strategies rely on fixed rules or pre-defined operating regimes, limiting adaptability to evolving attack strategies and dynamic grid conditions. Moreover, scalability and real-time responsiveness remain challenging in large, heterogeneous, and DER-rich distribution networks.

In response to these limitations, deep reinforcement learning (DRL) has emerged as a promising paradigm for enabling adaptive, scalable, and model-free robust voltage control in dynamic and uncertain distribution networks. An improved soft actor-critic (SAC) defense algorithm is developed in \cite{SelimAlaa10368040}, incorporating auto-tuned entropy, Gaussian policy modeling, and action prioritization, enabling the agent to allocate DER set points and load-shedding decisions robustly under cyber attacks. A robust attention-enhanced SAC method is proposed in \cite{YangXu10587051} for coordinated PV inverter control. The agent-level attention mechanism dynamically identifies the most influential agents for VVC, while a regularizer constrains the divergence between nominal and perturbed action distributions to enhance adversarial robustness. A robust deep deterministic policy gradient (R-DDPG) algorithm is proposed in \cite{ChenZhengrong11068158}, which integrates a min-max adversarial training framework to enhance resilience against adversarial perturbations. By incorporating prediction interval-based adversarial states that capture PV generation and load uncertainties, the agent learns robust policies under worst-case operating scenarios. In \cite{Roberts9482815}, a proximal policy optimization (PPO)-based DRL approach is developed to mitigate cyber attacks that manipulate DER Volt/Var and Volt/Watt setpoints, inducing voltage deviations across the network. The method coordinates uncompromised DERs by deploying an agent trained adaptively to adjust local droop parameters, thereby restoring voltage balance under adversarial conditions. In \cite{Wangyu10089185}, a twin delayed deep deterministic (TD3) DRL method is developed to model least effort attacks, which identify the minimum droop-gain perturbations required to destabilize the system, in inverter-based microgrids. A robust dynamic defense strategy is then generated through an adversarial DRL framework, in which the attacker and defender are formulated as adversarial Markov decision processes.

Despite these advances in DRL-based robust voltage control methods, existing research exhibits two key limitations. First, most studies evaluate robustness using randomly injected perturbations, ignoring system vulnerabilities and agent sensitivities. Such attacks lack strategic intent and fail to capture targeted adversarial behaviors that exploit grid dynamics or agent weaknesses. Second, prior research primarily focuses on black-box attacks, where the adversary has no knowledge of the control policy. In practice, white-box attacks, in which the attacker has full access to the DRL agent’s policy and gradient information, can be far more detrimental, enabling targeted, high-impact manipulations of system states and control actions. 

To address these gaps, this paper proposes a novel adversarial training framework for DRL-based VVO. Adversarial training using Projected Gradient Descent (PGD) is utilized to enable the agent to learn robust policies to high-impact, directionally optimized perturbations. As PGD is considered the strongest first-order adversary \cite{madry2019deeplearningmodelsresistant},  DRL agents trained against PGD-generated perturbations also exhibit strong robustness to weaker first-order attacks such as the fast gradient signed method (FGSM) \cite{goodfellow2015explainingharnessingadversarialexamples}, highlighting the effectiveness of gradient-based adversarial training in improving general robustness. 

\section{Problem Formulation}

\subsection{Power System Modeling for Distribution Networks}

Let the distribution network be represented by a graph $\mathcal{G} = (\mathcal{N}, \mathcal{E})$, where $\mathcal{N}$ denotes the set of buses and $\mathcal{E}$ denotes the set of branches. The complex voltages of all the buses in the distribution network, denoted by $U_i \in \mathbb{C}$ and $i \in \mathcal{N}$, are obtained by solving the nonlinear AC power flow equations, expressed as
\begin{align} \label{eq: PF_1}
     P^G_i - P^L_i & = \left|U_i\right|\sum_{j\in\mathcal{N}}\left|U_j\right|\left(G_{ij}\cos \theta_{ij} + B_{ij}\sin \theta_{ij}\right) \\ \label{eq: PF_2}
     Q^G_i - Q^L_i & = \left|U_i\right|\sum_{j\in\mathcal{N}}\left|U_j\right|\left(G_{ij}\sin \theta_{ij} - B_{ij}\cos \theta_{ij}\right) .
\end{align}
In \eqref{eq: PF_1} and \eqref{eq: PF_2}, $P^G_i$ and $Q_i^G$ denote the generated active and reactive power, and $P^L_i$ and $Q_i^L$ are the active and reactive load at bus $i$, respectively; $G_{ij}$ and $B_{ij}$ are the conductance and susceptance, and $\theta_{ij}$ is the voltage phase difference between buses $i$ and $j$, respectively.

Four types of voltage regulating devices are considered: smart inverters (SIs), CBs, OLTCs, and BESSs. For SIs,  it is assumed that they operate under the reactive power control mode, and the active power output of the distributed generations (DGs) is known. Then, the injected or absorbed reactive power of the SI on bus $i$ can be expressed as
\begin{align} \label{eq: QDG_1}
    &\underline{Q}_{DG, i} \leq Q_{DG, i} \leq \overline{Q}_{DG, i} \\ \label{eq: QDG_2}
    &\overline{Q}_{DG, i} = \sqrt{(S_{DG, i})^2 - (P_{DG, i})^2}.
\end{align}
In \eqref{eq: QDG_1} and \eqref{eq: QDG_2}, $\overline{Q}_{DG, i}$ and $\underline{Q}_{DG, i}$ denote the maximum and minimum available reactive power based on the DG’s apparent power $S_{DG, i}$ and active power $P_{DG, i}$. The control action of an SI on bus $i$ is defined as $a_{DG, i} \in [-1, 1]$, and $Q_{DG, i} = a_{DG, i} \, \overline{Q}_{DG, i}$. To facilitate the efficient and flexible training of DRL algorithms, the action space of SIs is discretized within its range with a step size of $0.1$ \cite{Zhang9143169}.

The CBs are modeled as follows:
\begin{align} 
    Q_{CB, i} = a_{CB, i}(U_{i})B_{i} \label{eq: CB}.
\end{align}
In \eqref{eq: CB}, $B_{i}$ denotes the susceptance of bus $i$ and $a_{CB, i} \in \{0, 1\}$ denotes the control action of CB on bus $i$, indicating the off/on status. 

The OLTCs are modeled as discrete voltage regulators with a regulating range of $\pm10\%$ with $32$ steps, expressed as
\begin{align} \label{eq: VR}
    &U_{VR, i} = 1\;\text{p.u}. + a_{VR, i}\Delta U_{VR}.
\end{align}
In \eqref{eq: VR}, $a_{VR, i} \in \{-16, -15, \dots, 16\}$ denotes the control action and the step for the regulator on bus $i$, and $\Delta U_{VR}=0.00625$ p.u. denotes the step size. 

The BESS at bus $i$ is modeled as a controllable active power source with state of charge (SoC) $\sigma_i$. Its SoC evolves according to
\begin{align} \label{eq: SoC}
    \sigma_i' = \sigma_i + \frac{P_i^\mathrm{bat} \Delta t}{E_i}.
\end{align}
In \eqref{eq: SoC}, $\Delta t$ denotes the time step and $E_i$ denotes the battery capacity. The SoC is constrained within operational limits
\begin{align}
    \sigma_{\min,i} \le \sigma_i' \le \sigma_{\max,i}.
\end{align}
The BESS control action, $a_{\mathrm{bat},i} \in [-1,1]$, determines the injected or absorbed power, where negative values indicate charging and positive values indicate discharging, as follows:
\begin{align}
    P_i^\mathrm{bat} = a_{\mathrm{bat},i} \, P_{\max,i}^\mathrm{bat}.
\end{align}
For DRL implementation, the continuous action is discretized into 33 levels to ensure fine-grained control while keeping the action space tractable, as follows:
\begin{align} \label{eq: bess_action_disc}
    a_{\mathrm{bat},i} \in \left\{-1, -1+d, \dots, 0, \dots, 1-d, 1\right\} \quad d=0.0625.
\end{align}
The objective of VVO is to regulate the bus voltage magnitudes within the standard operational range while minimizing total active power losses in the network. By solving \eqref{eq: PF_1} and \eqref{eq: PF_2}, the total active power loss in the system can be obtained by
\begin{align} \label{eq: Ploss}
    P_{\mathrm{loss}} = \sum_{(i,j) \in \mathcal{E}} \mathrm{Re}\left\{ {U}_{ij}^\top {I}_{ij} \right\}.
\end{align}
In \eqref{eq: Ploss}, ${U}_{ij} $ denotes the voltage drop across the branch $(i,j)$, and ${I}_{ij}$ denotes the corresponding line current obtained from the branch impedance.

The VVO problem can be formulated as an optimization problem with the following objective function:
\begin{equation} \label{eq: VVO_opt}
\begin{aligned}
\mathbf{a}
= \arg\min \;
& C_p P_{\mathrm{loss}}
  + C_v \sum_{i \in \mathcal{N}} \big( |U_i| - 1 \;\text{p.u.} \big)^2 \\
\text{s.t. } \quad
& \text{power flow equations } \eqref{eq: PF_1},\;\eqref{eq: PF_2}, \\
& \text{device constraints } \eqref{eq: QDG_1} -\eqref{eq: bess_action_disc}.
\end{aligned}
\end{equation}

In \eqref{eq: VVO_opt}, $C_p$ and $C_v$ denote the cost coefficients of power loss and voltage violation. While \eqref{eq: VVO_opt} only accounts for a single time step, in practice, sequential time-varying operating conditions are applied to the distribution network, which produces a sequence of VVO problems to be solved by the DRL algorithm.

\subsection{Markov Decision Process} \label{sec: MDP}
The VVO problem is formulated as a Markov Decision Process (MDP) defined by the tuple $(\mathcal{S}, \mathcal{A}, p, \mathcal{R})$ and over a finite horizon $H=24$, corresponding to one control action per hour in a $24$-hour daily cycle. The state space $\mathcal{S}$ is represented as
\begin{align} \label{eq: state_space}
    s^t = \big[ |\mathbf{U}^t|, \boldsymbol{\sigma}^{t-1}_\mathrm{BESS}, \mathbf{a}^{t-1}_\mathrm{VR},    \mathbf{a}^{t-1}_\mathrm{CB}, \mathbf{a}^{t-1}_\mathrm{DG} \big].
\end{align}
In \eqref{eq: state_space}, $|\mathbf{U}|$ denotes the vector of bus voltage magnitudes, $\boldsymbol{\sigma}_\mathrm{BESS}$ denotes the battery states of charge, and $\mathbf{a}_\mathrm{VR}$, $\mathbf{a}_\mathrm{CB}$, and $\mathbf{a}_\mathrm{DG}$ denote the discrete control actions of OLTCs, capacitor banks, and smart inverters, respectively. The action space $\mathcal{A}$ comprises setpoints for all controllable devices, denoted by the action vector $\mathbf{a}_t \in \mathcal{A}$. 

The transition probability $p(s^{t+1} \mid s^t, \mathbf{a}^t) : \mathcal{S} \times \mathcal{S} \times \mathcal{A} \rightarrow [0,1)$ represents the probability of reaching the next state $s_{t+1} \in \mathcal{S}'$ from the current state $s_{t} \in \mathcal{S}$ under action $\mathbf{a}_{t}$. Transitions are governed by the AC power flow equations \eqref{eq: PF_1}--\eqref{eq: PF_2} and device dynamics, and are therefore deterministic given the current state and action, except for stochastic variations in load and renewable generation.

The reward function is designed to align with the VVO objective described in \eqref{eq: VVO_opt}, integrating power loss minimization, voltage regulation, and device operation penalties into a unified scalar signal for reinforcement learning. Following the formulation in \cite{fan2022powergymreinforcementlearningenvironment}, the reward at each step is expressed as
\begin{align} \label{eq: reward}
    r(s^t, s^{t+1}, t) = & - f_\mathrm{volt}(s^{t+1}) - f_\mathrm{ctrl}(s^t, s^{t+1}, t) \nonumber \\
                         & - f_\mathrm{power}(s^{t+1}).
\end{align}
In \eqref{eq: reward}, $s_t$ and $s_{t+1}$ denote the system states before and after applying action $\mathbf{a}_t$ at time step $t$. The three penalty components respectively represent voltage deviation from nominal limits, control operation costs associated with inverters, capacitors, regulators, and batteries, and normalized total power loss. This formulation encourages the agent to maintain voltage profiles within acceptable bounds while minimizing active power losses and avoiding excessive device switching. Further implementation details of each component follow the definitions provided in \cite{fan2022powergymreinforcementlearningenvironment}.

Having defined the VVO problem as an MDP with states, actions, and rewards, a Double Deep Q-Network (DDQN) agent is employed to solve it. DDQN is selected because the VVC actions in PowerGym are inherently discrete, for which Q-learning methods are well-suited. Moreover, DDQN mitigates the overestimation bias present in standard DQN, providing improved training stability and more reliable value estimates in large combinatorial action spaces. The agent interacts with the environment by observing the current state $s_t$, selecting an action $\mathbf{a}_t$, receiving the reward $r_t$, and updating the state-action value estimates $Q(s_t, \mathbf{a}_t)$ via the DDQN framework.

\section{Proposed Adversarial Training for Robust Voltage Control}

DRL-based voltage control methods achieve effective real-time voltage regulation and loss minimization. However, they are highly vulnerable to cyberattacks, measurement errors, and extreme operating conditions. Such cyber-physical threats necessitate the development of a robust learning framework that enhances the resilience of DRL agents under these risks. To this end, this section proposes using adversarial training, a defense method that exposes the agent to intentionally perturbed states during learning to improve robustness against adversarial inputs in real-world operation.

\subsection{Adversarial State Markov Decision Process} \label{sec: ASMDP}

The VVO problem under adversarial observation is cast as an Adversarial State Markov Decision Process (ASMDP), extending the standard MDP described in Section \ref{sec: MDP} to account for perturbed states. For each true state \(s\in\mathcal{S}\), let \(\Xi(s)\subset\mathcal{S}\) denote the family of admissible adversarial states the adversary may generate. The set \(\Xi(s)\) constrains the adversary’s power, typically via an $\ell_{2}$ or $\ell_{\infty}$ norm bound on the perturbation. Then, the ASMDP can be represented by the tuple $(\mathcal{S},\mathcal{A},\Xi,r,p)$.

Under an adversary that presents \(s_{\mathrm{adv}}\in \Xi(s)\) to the agent, policy evaluation follows modified Bellman updates, which, for a DDQN agent, can be expressed as the following:
\begin{align} \label{eq: Bellman}
    \tilde{Q}(s,a) &= r(s,a,s') + \gamma \, \tilde{Q}\Big(s'_{\mathrm{adv}}, \arg\max_{a'} Q(s'_{\mathrm{adv}},a') \Big).
\end{align}
In \eqref{eq: Bellman}, \(\tilde{Q}(s,a)\) denotes the Q-value under adversarially perturbed states,  $s'_{\mathrm{adv}}\in\Xi(s')$ denotes the perturbed next state after the environment transition, and $\gamma\in[0,1]$ denotes the discount factor.  The target network provides a stable estimate of the maximal Q-value. This formulation ensures that the DDQN agent learns value functions and policies robust to adversarial perturbations while retaining the original MDP dynamics and reward structure.

\subsection{Gradient-Based Adversarial Attacks} \label{sec: Grad_based_attack}

\begin{algorithm}[!t]
\caption{Offline Adversarial Training with PGD for DDQN-Based Voltage Control}
\label{alg: adv_training_pgd}
\begin{algorithmic}[1]
\STATE \textbf{Input:} Replay buffer $\mathcal{D}$, Q-network $Q_\theta$, target network $Q_{\theta^-}$, perturbation radius $\delta$, PGD step size $\alpha$, PGD steps $K$, episode attack probability $p_{\mathrm{attack}}$, attack start episode $E_{\mathrm{attack}}$
\STATE \textbf{Initialize:} $Q_\theta$ randomly;  periodically update $Q_{\theta^-} \gets Q_\theta$
\FOR{each training episode}
    \FOR{each training iteration within the episode}
        \STATE Sample minibatch $\{(s,a,r,s')\}$ from $\mathcal{D}$
        \IF{episode $\ge E_{\mathrm{attack}}$, with probability $p_{\mathrm{attack}}$}
            \STATE Generate adversarial next state $s'_{\mathrm{adv}}$ using $K$ PGD steps within an $\ell_\infty$ ball of radius $\delta$
        \ELSE
            \STATE Set $s'_{\mathrm{adv}} \gets s'$
        \ENDIF
        \STATE Select action $a$ via $\epsilon$-greedy on current state $s$
        \STATE Store $(s, a, r, s'_{\mathrm{adv}})$ in $\mathcal{D}$
        \STATE Update $Q_\theta$ by minimizing DDQN loss using $Q_{\theta^-}$
        \STATE Periodically update target network: $Q_{\theta^-} \gets Q_\theta$
    \ENDFOR
\ENDFOR
\end{algorithmic}
\end{algorithm}

Adversarial attacks are modeled as additive perturbations $\eta$ applied to the state $s$, resulting in a perturbed observation
\begin{align} \label{eq: s_adv}
    s_{\mathrm{adv}} = s + \eta.
\end{align}
In \eqref{eq: s_adv}, $\eta$ is constrained by an $\ell_\infty$ norm such that $\|\eta\|_\infty \le \delta$. This formulation allows us to consider a range of first-order gradient-based attacks under a unified norm bound.

Among first-order attacks, the Fast Gradient Sign Method (FGSM) perturbs the state along the sign of the gradient of a loss function as follows \cite{goodfellow2015explainingharnessingadversarialexamples}:
\begin{align} \label{eq: FGSM}
    \eta_{\text{FGSM}} = \delta \cdot \text{sign}(\nabla_s J(s, \pi^{\text{target}})).
\end{align}
In \eqref{eq: FGSM}, $\delta$ denotes the maximum allowable perturbation magnitude, and  $J(s, \pi^{\mathrm{target}})$ denotes a surrogate loss function used to generate adversarial perturbations to measure the discrepancy between the current policy of the agent $\pi$ and a target distribution $\pi^{\mathrm{target}}$. Here, $\pi^{\mathrm{target}}$ is a hypothetical target policy constructed to emphasize the worst-case action   $a_{\mathrm{worst}} = \arg\min_{a \in \mathcal{A}} Q(s,a)$, which is the action that minimizes the Q-value at the current state. The target policy is used solely to define the loss for FGSM perturbation generation and is not used in actual DDQN value updates or action selection.

To generate stronger and more universal perturbations, Projected Gradient Descent (PGD) is utilized. PGD iteratively applies gradient-based updates while projecting the candidate states back onto the $\ell_\infty$ ball of radius $\delta$ around the original state. Specifically, the PGD update is expressed as
\begin{align} \label{eq: PGD_update}
    s^{(k+1)} = \Pi_{B_\infty(s, \delta)} \Big( s^{(k)} + \alpha \cdot \text{sign}(\nabla_s J(s^{(k)}, \pi^{\text{target}})) \Big).
\end{align}
In \eqref{eq: PGD_update}, $s^{(k)}$ denotes the candidate adversarial state at iteration $k$, where $k= 0,1,\dots,K-1$ with $K$ being the total number of PGD iterations; $\alpha$ denotes the PGD step size and $\Pi_{B_\infty(s, \delta)}(\cdot)$ denotes the projection operator that ensures the perturbation remains within the $\ell_\infty$ norm bound. PGD is considered a “universal” first-order adversary, providing robustness not only against itself but also against weaker gradient-based attacks like FGSM \cite{madry2019deeplearningmodelsresistant}. These perturbations define the set $\Xi(s)$ in the ASMDP formulation in Section~\ref{sec: ASMDP}, representing the family of admissible adversarial states the agent may encounter during training.

\subsection{Adversarial Training}

Adversarial training exposes the agent to worst-case PGD-perturbed states to enhance robustness against targeted cyberattacks. By probabilistically replacing nominal next states with PGD-generated adversarial states during DDQN updates, the agent learns value estimates and policies that maintain voltage regulation and minimize system losses under adversarial perturbations. Algorithm~\ref{alg: adv_training_pgd} illustrates this procedure and its integration within the standard offline DDQN training framework. Hyperparameters $\delta$, $\alpha$, $K$, and attack probability $p_{\text{attack}}$ are selected to reflect realistic bounds on adversarial perturbations, balancing robustness and training efficiency.

\begin{table}[!t]
\centering
\caption{\textsc{DDQN and Adversarial Attack Hyperparameters}}
\label{tab:combined_params}
\begin{tabular}{lc}
\hline
\multicolumn{2}{c}{DDQN Parameters} \\
\hline
Hidden layers, units, activation & 4, 512 each, ReLU \\
Optimizer, learning rate, $\gamma$ & Adam, $5\times10^{-5}$, 0.995 \\
Target network update frequency & 2000 steps \\
Replay buffer size, mini-batch size & 100,000, 512 \\
Exploration policy & \(\epsilon\)-greedy \\
Initial, final \(\epsilon\), \(\epsilon\)-decay & 1.0, 0.01, 0.9995 \\
Total episodes & 9,000 \\
\hline
\multicolumn{2}{c}{Shared Attack Parameters (PGD \& FGSM)} \\
\hline
Attack injection episode $E_{\mathrm{attack}}$ & 1000 \\
Attack scope & Voltage magnitude $|\mathbf{U}|$ \\
Attack probability per episode $p_{\mathrm{attack}}$ & 0.5 \\
\hline
\multicolumn{2}{c}{PGD Attack Parameters} \\
\hline
Step size (\(\alpha\)),  Max. perturbation (\(\delta\))& 0.0125, 0.1 \\
Number of iterations, random start & 20, Yes \\
\hline
\multicolumn{2}{c}{FGSM Attack Parameters} \\
\hline
Max. perturbation (\(\delta\)) & 0.1 \\
\hline
\end{tabular}
\vspace{-20pt}
\end{table}

During training, the experience tuple $(s, a, r, s'_{\mathrm{adv}})$ is stored in the replay buffer, and Q-network parameters are updated using the following DDQN loss:
\begin{align} \label{eq: DDQN_loss}
     \mathcal{L}(\theta) = \mathbb{E}_{(s, a, r, s'_{\mathrm{adv}})} \Big[ \big( & r + \gamma \max_{a'} Q(s'_{\mathrm{adv}}, a'; \theta^-) \nonumber \\
                                                                                 & - Q(s, a; \theta) \big)^2 \Big].
\end{align}
In \eqref{eq: DDQN_loss}, $\theta^-$ and  $\gamma$ denote the target network and the discount factor, respectively.

To prevent adversarial overfitting and instability observed when strong perturbations are introduced too early in training, PGD attacks are applied only with probability $p_{\text{attack}}$ and only after the agent has formed a stable baseline policy (e.g., episode $\geq E_{\text{attack}}$). This scheduling ensures that the agent first learns the nominal system dynamics before gradually learning to counter adversarial perturbations.

\begin{table}[!t]
\centering
\caption{\textsc{Effectiveness Comparison of FGSM \protect\\and PGD Methods}}\label{tab: perf_degrad}
\begin{tabular}{lccc}
\toprule
Metric                    & Clean & FGSM & PGD \\
\midrule
Reward                             &  -7.11 & -23.34 & -19.96 \\
Q-value reduction                  &   -- &   0.93 &   1.88 \\
Total voltage violations           &    650 &   3000 &   5437 \\
Avg. voltage violations/timestep    &   0.54 &   2.50 &   4.53 \\
\% at Max. $\delta$ (magnitude)      &   --  &  92.1  &  43.9  \\
\bottomrule
\end{tabular}
\vspace{-20pt}
\end{table}

By probabilistically incorporating PGD perturbations, the agent learns policies that are robust to adversarial attacks while maintaining nominal performance. This systematically strengthens the policy against input manipulations, aligning with a robust control perspective where adversarial perturbations represent worst-case deviations in observed states. Although PGD adversarial training increases computational cost and may slightly slow convergence, our case studies demonstrate that it significantly improves robustness, as PGD generates stronger adversarial perturbations than FGSM and effectively trains the agent against worst-case attacks.

\section{Case Study}

The proposed robust control method is tested on the IEEE 123-bus distribution network to validate its effectiveness. A total of $15$ baseline load profiles, provided by Powergym \cite{fan2022powergymreinforcementlearningenvironment}, is utilized and split into $10$ for training, $2$ for evaluation, and $3$ for testing. The baseline load profiles are perturbed between $80-120\%$ uniformly to generate diverse system loading scenarios. Twenty photovoltaic systems, interfaced with the grid through SIs, are installed on buses 7, 12, 18, 22, 25, 37, 44, 52, 64, 77, 81, 85, 90, 91, 92, 100, 103, 106, 111, and 116. In addition, CBs, OLTCs, and BESSs are installed on buses and branches as defined in the Powergym environment. Each simulation episode consists of $24$ time steps.

Table~\ref{tab:combined_params} summarizes the DDQN training parameters and adversarial attack settings used in this study. Three models are trained: a baseline model without adversarial attacks, a model trained with FGSM-based attacks, and a model trained with PGD-based attacks. Perturbation magnitude ($\delta = 0.1$) is chosen to challenge the agent without causing unrealistic states. Step size and number of iterations are chosen to ensure that the boundary of the $\delta$-ball can be reached from any starting point within it \cite{madry2019deeplearningmodelsresistant}. 

\begin{table}[!t]
\centering
\caption{\textsc{Cross-Attack Performance of Robustly Trained DDQN Agents}}
\label{tab:cross_attack_metrics_compact}
\begin{tabular}{l c c c c c}
\toprule
\multirow{2}{*}{\makecell{Train \\ Attack}} &
\multirow{2}{*}{\makecell{Test \\ Attack}} &
\multicolumn{2}{c}{Reward} &
\multirow{2}{*}{\makecell{Volt. Viol.\\ (Total)}} &
\multirow{2}{*}{\makecell{Total Switches \\ (per timestep)}} 
\\
\cline{3-4}
 & & Clean & Attack & & \\
\hline
\multirow{2}{*}{\makecell[c]{FGSM}}
    & FGSM & \multirow{2}{*}{-6.88} & -9.43  & 0   & 7.58\\
    & PGD  &                        & -10.35 & 257 & 7.18\\
\hline
\multirow{2}{*}{\makecell[c]{PGD}}
    & FGSM & \multirow{2}{*}{-6.73} & -9.16 & 0   & 8.69\\
    & PGD  &                        & -8.49 & 0   & 6.24\\
\bottomrule
\end{tabular}%
\vspace{-20pt}
\end{table}

\subsection{Effectiveness of Adversarial Attacks on Naively Trained Models}

Table~\ref{tab: perf_degrad} summarizes the impact of FGSM and PGD attacks on a standard, non-robust DDQN agent on the test dataset over $50$ episodes. The attack parameters are kept the same as during the training to ensure a fair comparison. For clarity, we define the metrics reported in the table. Reward is the mean episodic value across all episodes. Q-value reduction measures the average decrease in action value at each attack timestep. Voltage violations are counted per timestep if bus voltages exceed $\pm 5\%$ of nominal ($1.0$ p.u.), and the \% at maximum $\delta$ indicates how many perturbed measurements reach the maximum allowed magnitude.

Both gradient-based methods cause significant degradation, with the FGSM and PGD methods producing a $228\%$ and $181\%$ drop in mean reward compared to the clean baseline, respectively. The Q-value reduction provides insight into how much the attack degrades the estimated value of the actions chosen by the agent,  quantifying how severely the perturbations distort the learned value landscape. PGD causes a near two-fold larger reduction than the FGSM method, reflecting the effectiveness of its iterative optimization at steering the agent toward low-value regions of the state space. This effectiveness directly transfers to the number of voltage violations. 

Although FGSM produces a larger degradation in mean reward, the PGD attack generates substantially more voltage violations. This difference arises from the nature of the attacks. FGSM applies a single-step attack in which more than $92\%$ of the perturbed voltage measurements reach the maximum allowed magnitude. PGD, however, uses multiple gradient-based refinements within each attack timestep and produces perturbations that reduce Q-values more effectively, even though the perturbation magnitude is less than half of that used by FGSM. This suggests that PGD updates are better aligned with the agent’s value function, steering the policy toward actions that ultimately yield more unsafe voltage profiles over time.

\subsection{Adversarial Training and Robustness Evaluation}

\begin{figure} [!t] 
    \centering
    \includegraphics[width=\linewidth]{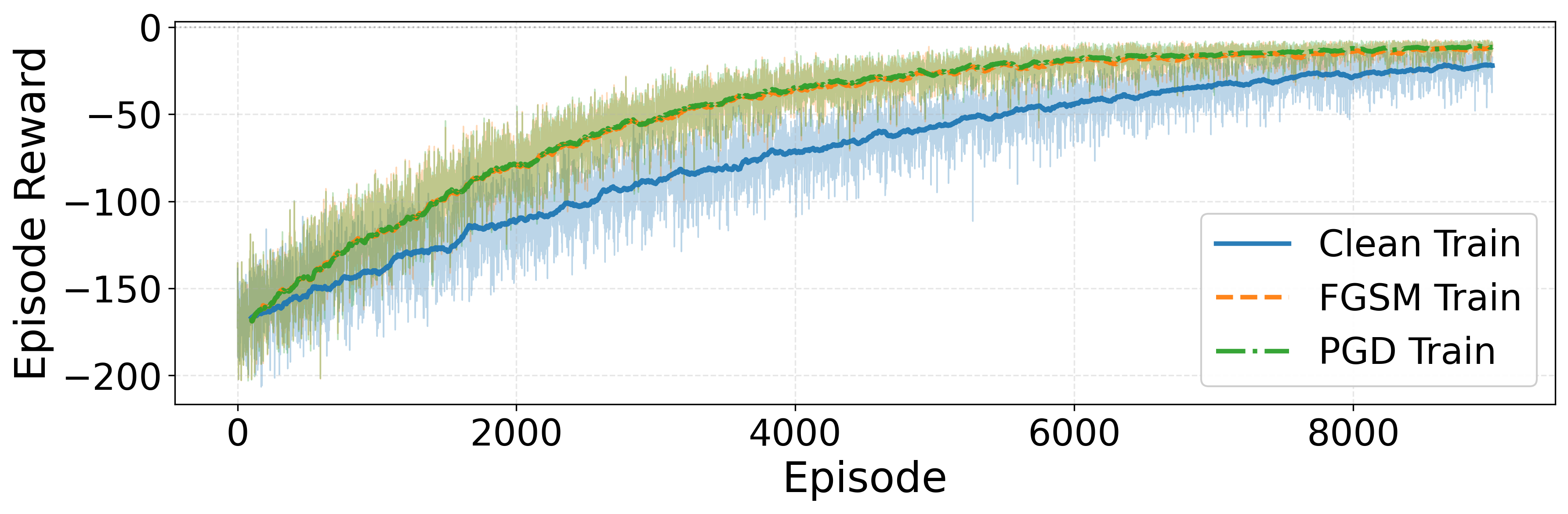}
    \caption{Episode reward curves during training for clean-, FGSM-, and PGD-trained DDQN agents}
    \label{fig: training_curve}
    \vspace{-10pt}
\end{figure}

Fig. \ref{fig: training_curve} illustrates the episode reward curves during training for clean-, FGSM-, and PGD-trained DDQN agents. Surprisingly, the clean-trained agent exhibits slower learning and achieves the lowest episodic reward, suggesting that training solely on nominal states limits exposure to challenging operating conditions. In contrast, adversarial training with FGSM or PGD not only improves robustness but also enhances exploration, guiding the agent to learn policies that generalize more effectively across the full state space.

\begin{figure}[t]
  \centering

  \subfloat[FGSM attack\label{fig:voltage_pgd}]{
    \includegraphics[width=\columnwidth]{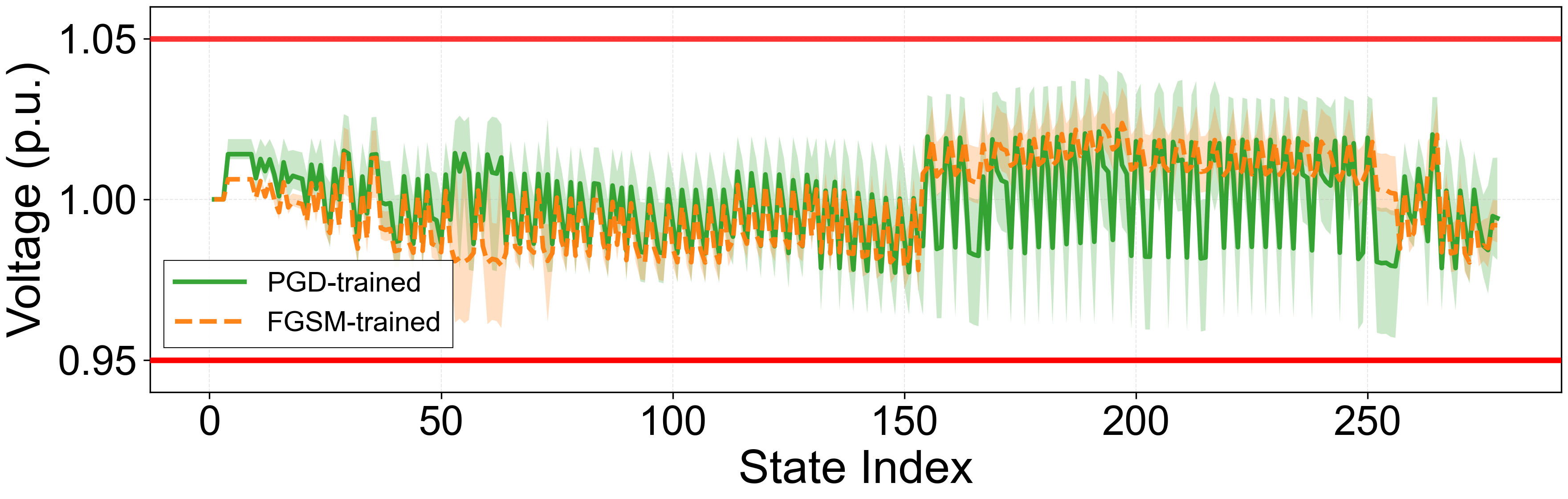}
  }\\[4pt]

  \subfloat[PGD attack\label{fig:voltage_fgsm}]{
    \includegraphics[width=\columnwidth]{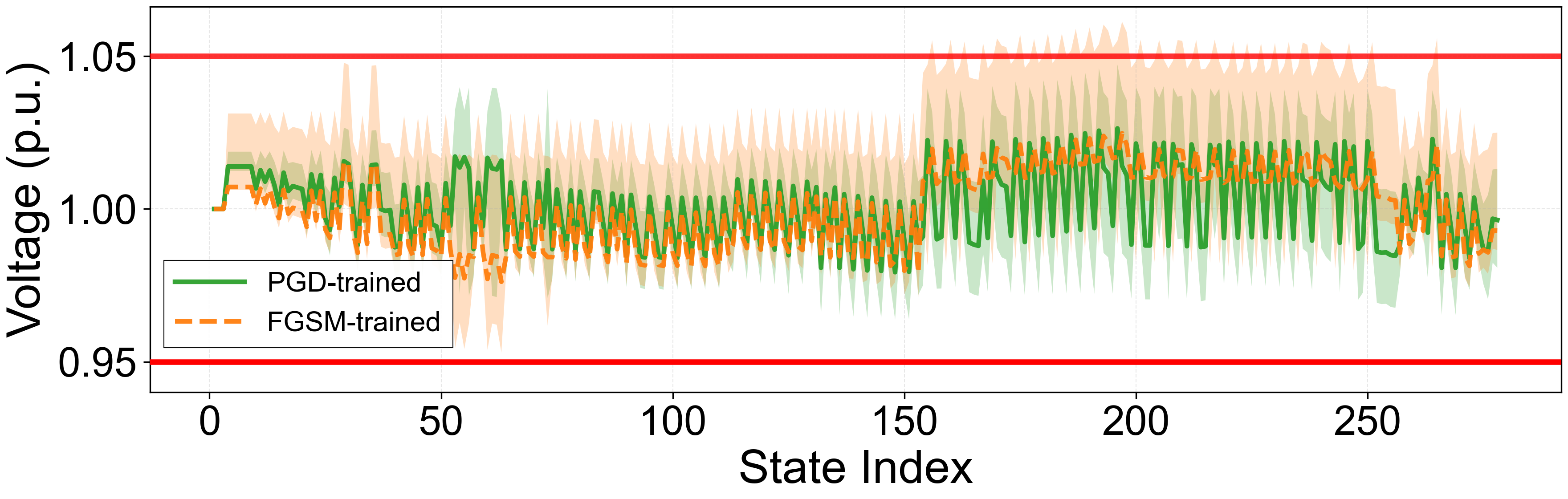}
  }

  \caption{Voltage profiles under different adversarial attacks.
  Shaded regions indicate the minimum--maximum voltage envelope across all episodes,
  solid curves represent the mean voltage, and horizontal red lines denote the
  operational limits (0.95--1.05~p.u.).}
  \label{fig:voltage_comparison_attacks}
  \vspace{-5pt}
\end{figure}

Table~\ref{tab:cross_attack_metrics_compact} summarizes the cross-attack performance of DDQN agents robustly trained against either FGSM or PGD attacks. Each model was evaluated on the test dataset for $50$ episodes under clean conditions and under both FGSM and PGD attacks to assess robustness generalization across attack types. When subjected to cross-type attacks, the PGD-trained agent effectively mitigates voltage violations while performing approximately $40\%$ more switches per timestep under cross-type attacks. In contrast, the FGSM-trained agent makes  $5\%$ fewer switches per timestep under cross-type attacks, failing to compensate for the stronger perturbations. The Q-value reductions further indicate that PGD-trained agents preserve their estimated action values roughly two times more effectively than FGSM-trained agents under cross-attacks, highlighting a marked improvement in policy stability. These findings are further reinforced by Fig. \ref{fig:voltage_comparison_attacks}, in which the FGSM-trained model exhibits a substantial increase in voltage range and number of voltage violations when subject to PGD attacks. Collectively, these observations demonstrate that PGD-based adversarial training not only sustains reward performance but also substantially enhances robustness and ensures constraint satisfaction in previously unseen attack scenarios.

\section{Conclusion}

This paper develops a DRL-based voltage control framework that leverages PGD-based adversarial training to enhance robustness against strategic cyber attacks. By exposing the agent to high-impact, directionally optimized perturbations, the trained policy maintains voltage stability and operational efficiency under realistic attack scenarios. Simulations show that PGD-trained agents also generalize well to weaker gradient-based attacks such as FGSM, highlighting the effectiveness of gradient-based adversarial training in improving resilience and adaptability in DER-rich distribution networks.

\IEEEpubidadjcol
\bibliographystyle{IEEEtran}
\bibliography{IEEEabrv,Citation}
\let\mybibitem\bibitem

\end{document}